\documentclass[a4paper,11pt]{article}
\pdfoutput=1 

\usepackage{jcappub} 
\usepackage{indentfirst}
\usepackage{appendix}
\usepackage{multirow}

\usepackage{array}
\usepackage{enumerate}
\usepackage{graphicx}
\usepackage{dcolumn}
\usepackage{CJKutf8}
\usepackage{bm}
\usepackage{amssymb}
\usepackage{latexsym}
\usepackage{booktabs}
\usepackage{amsmath}
\usepackage{subcaption}
\usepackage{url}
\newcommand{\ksm}{~{\rm km}~{\rm s}^{-1}~{\rm Mpc}^{-1}}
\begin{document}
\title{Enhancing dark siren cosmology through multi-band gravitational wave synergetic observations}

\author[a,1]{Yue-Yan Dong,\note{These authors contributed equally to this paper.}}
\author[a,1]{Ji-Yu Song,}
\author[a,b]{Shang-Jie Jin,}
\author[a]{Jing-Fei Zhang}
\author[a,c,d,2]{and Xin Zhang\note{Corresponding author.}}
\affiliation[a]{Key Laboratory of Cosmology and Astrophysics (Liaoning Province) and College of Sciences, Northeastern University, Shenyang 110819, China}
\affiliation[b]{Department of Physics, University of Western Australia, Perth WA 6009, Australia}
\affiliation[c]{Key Laboratory of Data Analytics and Optimization for Smart Industry (Ministry of Education), Northeastern University, Shenyang 110819, China}
\affiliation[d]{National Frontiers Science Center for Industrial Intelligence and Systems Optimization, Northeastern University, Shenyang 110819, China}

\emailAdd{dongyueyan@stumail.neu.edu.cn, songjiyu@stumail.neu.edu.cn, jinshangjie@stumail.neu.edu.cn, jfzhang@mail.neu.edu.cn, zhangxin@mail.neu.edu.cn}

\abstract{Multi-band gravitational-wave (GW) standard siren observations are poised to herald a new era in the study of cosmic evolution. These observations offer higher signal-to-noise ratios and improved localizations compared to those achieved with single-band GW detection, which are crucial for the cosmological applications of dark sirens. In this work, we explore the role multi-band GW synergetic observations will play in measuring cosmological parameters, particularly in comparison with single GW observatory data. We used mock multi-band dark siren data from third-generation GW detectors and the baseline Decihertz Interferometer Gravitational-Wave Observatory to infer cosmological parameters. Our analysis shows that multi-band GW observations significantly improve sky localization accuracy by two to three orders of magnitude over single-band observations, although their impact on luminosity distance error remains limited. This results in a substantial improvement in the constraints on matter density and the Hubble constant, enhancing their constraint precision by $60\%$--$90\%$ and $52\%$--$85\%$, respectively. We conclude that the significant potential of multi-band GW synergistic observations for detecting GW signals and resolving the Hubble tension is highly promising and warrants anticipation.}

\maketitle
\section{Introduction}\label{sec:intro}

The Hubble tension has emerged as a significant enigma in cosmology in recent years \cite{Verde:2019ivm,Riess:2019qba,DiValentino:2021izs,Perivolaropoulos:2021jda,Abdalla:2022yfr,Kamionkowski:2022pkx}. This confusion stems from a more-than-5$\sigma$ descrepency between the values of the Hubble constant ($H_0$) inferred from the Planck 2018 CMB observation \cite{Planck:2018vyg} based on the $\Lambda$CDM model (a 0.8\% measurement) and obtained through the model-independent distance ladder measurement (a 1.4\% measurement) \cite{Riess:2021jrx}. This inconsistency hints at the possibility of new physics beyond the $\Lambda$CDM model (see, e.g., refs.~\cite{Li:2010xjz,Li:2013dha,Zhang:2014nta,Zhang:2014ifa,Zhang:2014dxk,Zhao:2017urm,Feng:2017nss,Yang:2018euj,Poulin:2018cxd,Guo:2018ans,DiValentino:2019ffd,DiValentino:2019jae,Liu:2019awo,Zhang:2019cww,Ding:2019mmw,Vagnozzi:2019ezj,Guo:2019dui,Feng:2019jqa,Li:2020tds,Lin:2020jcb,Mukherjee:2020hyn,DiValentino:2020zio,Gao:2021xnk,Cai:2021wgv,Vagnozzi:2021gjh,Vagnozzi:2021tjv,Wang:2021kxc,Gao:2022ahg,Zhao:2022yiv,Qi:2024acx,Du:2024pai,Li:2024qso}). However, no consensus is reached on a valid extended cosmological model that can truly resolve the Hubble tension \cite{Guo:2018ans}. Therefore, one may pay more attention to some promising and powerful cosmological probes that can independently measure $H_0$ to help resolve the Hubble tension. The gravitation wave (GW) standard sirens are a promising new cosmological probe that could help determine the $H_0$ value independently, which is widely discussed in refs.~\cite{Dalal:2006qt,Cutler:2009qv,Nissanke:2009kt,Zhao:2010sz,Cai:2016sby,Cai:2017cbj,Cai:2017aea,Cai:2017plb,Zhang:2018byx,Du:2018tia,Zhang:2019ylr,Belgacem:2019tbw,Safarzadeh:2019pis,Zhang:2019loq,Chang:2019xcb,Zhang:2019ple,Yang:2019bpr,Bachega:2019fki,He:2019dhl,Zhao:2019gyk,Wang:2019tto,Chen:2020dyt,Jin:2020hmc,Borhanian:2020vyr,Hogg:2020ktc,Nunes:2020rmr,Chen:2020zoq,Mitra:2020vzq,Qi:2021iic,Fu:2021huc,Bian:2021ini,Ye:2021klk,Jin:2021pcv,Guo2021StandardSC,Yu:2021nvx,Cao:2021zpf,deSouza:2021xtg,Zhu:2022dfq,Wu:2022dgy,Jin:2022tdf,Wang:2022oou,Dhani:2022ulg,Colgain:2022xcz,Hou:2022rvk,Califano:2022syd,Jin:2022qnj,Jin:2023zhi,Han:2023exn,Wang:2023lif,Jin:2023tou,Vagnozzi:2023nrq,Li:2023gtu,Yang:2023zxk,Zheng:2024mbo,Feng:2024mfx} and references therein.

Compact binary coalescence (CBC) events produce GWs whose waveforms encode the luminosity distance information. Through the GW waveform analysis, one can directly obtain the luminosity distance, which is vividly referred to as a standard siren \cite{Schutz:1986gp,Holz:2005df}. Once the redshifts of GW sources are determined, the established distance-redshift relation can be used to explore the history of cosmic expansion. There are two main methods regarding the redshift determinations of GW sources adopted by the current GW observations. One is to determine redshift through its electromagnetic (EM) counterparts, an event that relies on the binary neutron star (BNS) or neutron star-black hole (NSBH) merger (usually referred to as a bright siren). The other is to use a galaxy catalog, inferring the redshift through a statistical method (referred to as a dark siren) \cite{DelPozzo:2011vcw,Chen:2017rfc,Nair:2018ign,LIGOScientific:2018gmd,DES:2019ccw,Gray:2019ksv,DES:2020nay,Yu:2020vyy,LIGOScientific:2019zcs,Finke:2021aom,Leandro:2021qlc,Liu:2021yoy,Yang:2022iwn,Song:2022siz,LIGOScientific:2021aug,Palmese:2021mjm,Zhu:2021aat,Gair:2022zsa,Mukherjee:2022afz,Yang:2021xox,Yang:2022tig,Muttoni:2023prw,Li:2023gtu,Zhu:2023jti,Yu:2023ico,Zhu:2024qpp,Xiao:2024nmi}.

So far, the first, second, and third observation runs from the LIGO-Virgo-KAGRA (LVK) collaboration have reported over 90 CBC events \cite{LIGOScientific:2018mvr,LIGOScientific:2020ibl,LIGOScientific:2021djp}. However, owing to the challenge in detecting EM counterparts, only one bright siren, GW170817, has been identified, which yields approximately a 14\% constraint on $H_0$ \cite{LIGOScientific:2017adf}. For dark siren analysis, using 46 dark siren events with SNRs over 11, combined with the GLADE+ catalog \cite{Dalya:2018cnd,Dalya:2021ewn}, achieves a 19\% constraint on $H_0$ \cite{KAGRA:2021vkt,LIGOScientific:2021aug}. Additionally, the 46 dark sirens combined with the one bright siren GW170817 constrain $H_0$ to around 10\% \cite{LIGOScientific:2021aug}, which has not yet reached the level required to resolve the Hubble tension. 

In the future, the multi-band relay detection of a GW event becomes possible. In the few hertz to kilohertz band, the cutting-edge third-generation (3G) ground-based GW detectors, i.e., the Einstein Telescope (ET) \cite{Punturo:2010zz} and the Cosmic Explorer (CE) \cite{LIGOScientific:2016wof}, are set to debut, anticipated to achieve sensitivity enhancements by an order of magnitude than current GW detectors. The space-based baseline Decihertz Interferometer Gravitational-Wave Observatory (B-DECIGO) \cite{10.1093/ptep/ptw127} with longer arms will provide a detection window in the decihertz band. In addition, GW signals in the millihertz band will be detected by space-based GW detectors with even longer arms, such as LISA \cite{LISA:2017pwj,Robson:2018ifk,LISACosmologyWorkingGroup:2022jok}, Taiji \cite{Hu:2017mde,Wu:2018clg,Ruan:2018tsw}, and TianQin \cite{TianQin:2015yph,Wang:2019ryf,Liu:2020eko,Luo:2020bls,Milyukov:2020kyg,TianQin:2020hid}. The concept of the multi-band GW observations has been widely investigated in refs.~\cite{Sesana:2016ljz,Vitale:2016rfr,Sesana:2017vsj,Isoyama:2018rjb,Jani:2019ffg,Carson:2019kkh,Liu:2020nwz,Datta:2020vcj,Zhang:2021pwe,Nakano:2021bbw,Yang:2021qge,Muttoni:2021veo,Liu:2021dcr,Zhu:2021bpp,Kang:2022nmz,Klein:2022rbf,Seymour:2022teq,Baker:2022eiz,Zhao:2023ilw}, and several advantages are dug out: providing early warnings of the detection of BNSs, which is crucial of EM counterparts searches; enhancing the accumulation of signal-to-noise (SNR); improving the precisions of GW parameters inference, including the spatial localization precision.

We notice the current dark sirens analysis with GWTC-3 and GLADE+ catalog has very poor localization, with sky localization primarily a few hundred square degrees, occasionally extending to a few thousand, and high luminosity distance measurement errors, ranging from 40 Mpc to several thousands of Mpc at the 1$\sigma$ level. We anticipate that in the future, using the multi-band synergetic detection of the GW events will substantially enhance spatial localization precision, thereby improving the precision of constraints on cosmological parameters with dark sirens. Therefore, we wish to study how multi-band synergetic detection of GW events enhances the dark siren cosmology.

In this paper, we simulate the multi-band synergetic detection of B-DECIGO and 3G ground-based GW detectors for the stellar-mass binary black hole (SBBH) GW events. We use the mock galaxy catalog to provide redshift information for GW dark sirens and employ a Bayesian analysis approach to infer cosmological parameters. By changing the apparent magnitude threshold to simulate galaxy catalogs with different completeness, we analyze the influence of the completeness of the galaxy on our results. By comparing the constraint results of cosmological parameters from multi-band and single-band observations, we explore the role that future multi-band observations will play in the GW dark siren cosmology.

This paper is organized as follows. In section~\ref{Method}, we introduce the methods of simulating the multi-band synergetic detection GW data and the dark siren analysis. In section~\ref{sec:results and discussion}, we give the constraint results and make detailed discussions. The conclusion is given in section~\ref{Conclusion}.

\section{Method}\label{Method}

\subsection{Cosmological model}\label{Cosmological models}
In the expanding universe, the luminosity distance $d_{\rm L}$ can be expressed theoretically as
\begin{equation}\label{eq:d-z relation}
d_{\rm L}(z)=c(1 + z)\int_{0}^{z}\frac{{\rm d}z^{\prime}}{H(z^{\prime})},
\end{equation}
where $c$ is the speed of light in vacuum, and $H(z)$ is the Hubble parameter, describing the universe's expansion rate at redshift $z$.

In the standard model of cosmology, the $\Lambda$CDM model, the equation of state parameter of dark energy $w=-1$, and $H(z)$ can be expressed as
\begin{equation}
H(z) = H_0\sqrt{\Omega_{\rm m}(1 + z)^3 + 1 - \Omega_{\rm m}}.
\end{equation}
In our GW data simulation, we adopt the $\Lambda$CDM model as the fiducial model with $H_0=67.27\ {\rm km~s^{-1}~Mpc^{-1}}$ and $\Omega_{\rm m}=0.3166$ from the constraint results of Planck 2018 TT, TE, EE+lowE \cite{Planck:2018vyg}.

\subsection{Simulation of GW sources}\label{sec:simulate GW source}
In this study, we simulate the detection of SBBHs by future GW detectors. The number of SBBHs within the redshift range from $z_{\rm L}$ to $z_{\rm U}$ is given by
\begin{equation}
N_{\rm GW} = \int_{z_{\rm L}}^{z_{\rm U}} R_{\rm obs}(z) {\rm d}z,
\end{equation}
where $R_{\rm obs}(z)$ is the SBBH merger rate at $z$ in the observer frame, which can be further converted to the source-frame merger rate $R_{\rm m}(z)$ via
\begin{equation}\label{eq:Robs}
    R_{\rm obs}(z) = \frac{R_{\rm m}(z)}{1 + z}\frac{{\rm d}V_{\rm c}(z)}{{\rm d}z}, 
\end{equation}
where $V_{\rm c}(z)=4\pi/3d^3_{c}(z)$ is the comoving volume. $R_{\rm m}(z)$ is related to the formation rate of binary systems through the time delay distribution,
\begin{equation}
    R_{\rm m}(z_{\rm m}) = \int_{z_{\rm m}}^\infty {\rm d}z_{\rm f}\frac{{\rm d}t_{\rm f}}{{\rm d}z_{\rm f}} R_{\rm f}(z_{\rm f}) P(t_{\rm d}),
\end{equation}
where $R_{\rm f}(z)$ is the formation rate of the binary system, assumed to be proportional to the Madau-Dickinson (MD) star formation rate \cite{Madau:2014bja},
\begin{equation}
    R_{\rm f}(z) = A \left(1 + z \right)^{2.7} \frac{1}{1 + \left[( 1 + z )/2.9\right]^{5.6}},
\end{equation}
where $A$ is the normalization factor, determined by the BBH merger rate at $z = 0$, and we set $R_{\rm m}(0) = 23.9$ Gpc$^{-3}$ yr$^{-1}$, which is the estimated median rate of O3 observation \cite{KAGRA:2021duu}.

$P(t_{\rm d})$ is the distribution of the time delay, and we adopt the exponential form as \cite{Vitale:2018yhm}
\begin{equation}
    P(t_{\rm d}) = \frac{1}{\tau} \exp \left( -\frac{t_{\rm d}}{\tau} \right), 
\end{equation}
where $t_{\rm d} = t_{\rm f} - t_{\rm m}$ is the time delay, with $t_{\rm m}$ being the merger time, which also corresponds to the look-back time at the redshift $z_{\rm m}$. The parameter $\tau$ is set to 100 Myr.

We obtain the masses of SBBHs based on the \textsc{power law + peak} model \cite{LIGOScientific:2020kqk,KAGRA:2021duu}, which provides a better fit to the observed mass distribution by capturing both the overall power-law trend and the excess around 30--40\,$M_\odot$, expressed as
\begin{equation}
p(m_1) = \left[(1 - \lambda_{\rm peak}) B(m_1) + \lambda_{\rm peak} G(m_1)\right] S(m_1),
\end{equation}
where $m_1$ ranges from 5 to 44 $M_\odot$, and $\lambda_{\rm peak} = 0.038$. $B(m_1) \propto m_1^{-\alpha}$ represents a normalized power-law distribution with the spectral index $\alpha = 3.5$, while $G(m_1)$ is a Gaussian distribution with the mean value $\mu_m = 34~M_\odot$. $S(m_1)$ is a smoothing function, which is given by
\begin{equation}
    S(m_1) = 
    \begin{cases} 
        0, & {\rm if ~} m_1 < m_{\rm min}, \\
        \frac{1}{f \left(m_1 - m_{\rm min}\right) + 1}, & {\rm if~} m_{\rm min} \leq m_1 < m_{\rm min} + \delta_m, \\
        1, & {\rm if~} m_{\rm min} + \delta_m \leq m_1,
    \end{cases}
\end{equation}
with
\begin{equation}
    f(m) = \exp \left(\frac{\delta_m}{m} + \frac{\delta_m}{m - \delta_m}\right), \qquad \delta_m = 4.9\, M_\odot.
\end{equation}
The probability distribution of the mass ratio $q$ can be expressed as
\begin{equation}
    p(q) \propto q^\beta S(m_1 q),
\end{equation}
with $\beta = 1.1$. The secondary mass of the BBH is determined as a function of the mass ratio $q$. The remaining parameters, including colatitude $\theta$, longitude $\phi$, inclination angle $\iota$, polarization angle $\psi$, coalescence phase $\varphi_{\rm c}$, and coalescence time $t_{\rm c}$ are randomly sampled within the ranges $\cos\theta \in [-1, 1]$, $\phi \in [0, 2\pi)$, $\cos\iota \in [-1, 1]$, $\psi \in [0, 2\pi)$, $\varphi_{\rm c} \in [0, 2\pi)$, and $t_{\rm c} \in [0, 4]~\mathrm{yr}$, respectively.

\subsection{Simulation of GW signals}
For a GW detector network composed of $N$ GW detectors, the Fourier transform of the time-domain waveform is given by \cite{Wen:2010cr,Zhao:2017cbb}
\begin{equation}\label{eq3}
\bm{\tilde{h}}(f) = e^{-i\Phi}\bm{\hat{h}}(f),
\end{equation}
where $\Phi$ is a $N\times N$ diagonal matrix, with $\Phi_{kl} = 2\pi f\delta_{kl}(\bm{n}\cdot\bm{r}_{k})$. Here, $\delta_{kl}$ is the Kronecker delta, $\bm{n}$ is the propagation direction of a GW, and $\bm{r}_{k}$ represents the location of $k$th GW detector. $\bm{\hat{h}}(f)$ is given by
\begin{equation}
\bm{\hat{h}}(f) = \big[\tilde{h}_{1}(f), \tilde{h}_{2}(f), \cdots , \tilde{h}_{N}(f)\big],
\end{equation}
where $\tilde{h}_k(f)$ is the frequency-domain GW waveform on the $k$th GW detector, given by 
\begin{align}\label{eq4}
\tilde{h}_{k}(f)= h_{+}(f)F_{+,k}(f) + h_{\times}(f)F_{\times,k}(f),
\end{align}
where $h_{+}(f)$ and $h_{\times}(f)$ are the waveform of the GW signal. In this work, we use {\tt Pycbc} \cite{Usman:2015kfa,Nitz:2017svb,Davies:2020tsx} and the phenomenological inspiral-merger-ringdown (IMR) GW model, IMRPhenomD \cite{Husa:2015iqa,Khan:2015jqa}, to simulate GW signals.
$F_{+,k}(f)$ and $F_{\times,k}(f)$ are antenna pattern functions of the $k$th GW detector, which are related to both the locations of the GW source and the detector. In this work, we analyze ET, two CEs, and B-DECIGO. CE in the USA has 40 km length of arms, denoted as CE (40 km), while the other in Australia with similar design but shortened 20 km length of arms, denoted as CE (20 km). The pattern functions of the L-shaped CE are given by \cite{Sathyaprakash:2009xs}
\begin{align}
F_{+}(\theta,\phi,\psi) = \frac{1}{2}(1+\cos^2{\theta})\cos{2\phi}\cos{2\psi} -\cos{\theta}\sin{2\phi}\sin{2\psi},\nonumber \\  
F_{\times}( \theta,\phi,\psi) = \frac{1}{2}(1+\cos^2{\theta})\cos{2\phi}\sin{2\psi}+\cos{\theta}\sin{2\phi}\cos{2\psi}.\label{eq:ce}
\end{align}

ET is a triangle GW detector with three 10-km arms, which could be equivalent to three interferometers with a 60$^\circ$ angle between each other, and the pattern functions of one of the interferometers are given by \cite{Punturo:2010zz}
\begin{align}
F_{+}^{(1)}( \theta,\phi,\psi) = \frac{\sqrt{3}}{2}[\frac{1}{2}(1+\cos^2{\theta})\cos{2\phi}\cos{2\psi} -\cos{\theta}\sin{2\phi}\sin{2\psi}], \nonumber \\ 
F_{\times}^{(1)}( \theta,\phi,\psi) = \frac{\sqrt{3}}{2}[\frac{1}{2}(1+\cos^2{\theta})\cos{2\phi}\sin{2\psi} 
+\cos{\theta}\sin{2\phi}\cos{2\psi}].
\end{align}
The other two interferometers's pattern functions are $F_{+,\times}^{(2)}( \theta,\phi,\psi)=F_{+,\times}^{(1)}( \theta,\phi+2\pi/3,\psi)$, and $F_{+,\times}^{(3)}( \theta,\phi,\psi)=F_{+,\times}^{(1)}( \theta,\phi+4\pi/3,\psi)$, respectively.
The antenna pattern functions of one of the interferometers of B-DECIGO are given by \cite{Yagi:2011wg,Kawamura:2020pcg}
\begin{align}
F_{+}^{(1)}( \theta,\phi,\psi) = \frac{1}{2}(1+\cos^2{\theta})\cos{2\phi}\cos{2\psi} -\cos{\theta}\sin{2\phi}\sin{2\psi},\nonumber \\  
F_{\times}^{(1)}( \theta,\phi,\psi) = \frac{1}{2}(1+\cos^2{\theta})\cos{2\phi}\sin{2\psi}+\cos{\theta}\sin{2\phi}\cos{2\psi},\label{eq:ce}
\end{align}
and the other interferometer's pattern functions are $F_{+,\times}^{(2)}( \theta,\phi,\psi)=F_{+,\times}^{(1)}( \theta,\phi+\pi/4,\psi)$. 
In the antenna pattern functions $F_{+}$ and $F_{\times}$, the angular parameters $\theta$, $\phi$, and $\psi$ evolve with the observation time $t$. Since our calculations employ waveforms in the frequency domain, we account for this variation by expressing the observation time as a function of frequency, given by $t(f) = t_c - 5(8\pi f)^{-\frac{8}{3}}{M_c}^{-\frac{5}{3}}$ \cite{Krolak:1995md, Buonanno:2009zt}. The explicit form of the angular parameter evolution can be found in ref.~\cite{Liu:2020nwz}.

The specific locations of ground-based GW detectors are referenced in refs.~\cite{10.1785/0220200186,Borhanian:2020ypi}, while the design of B-DECIGO can be found in ref.~\cite{Kawamura:2020pcg}. For more details on each GW detector's antenna pattern functions, refer to refs.~\cite{Jaranowski:1998qm,Arnaud:2001my,Kawamura:2006up,Schutz:2011tw,Yagi:2011wg}.
In addition, we include various network configurations. These configurations encompass the 2CE network, consisting of two CEs; the ET2CE network, comprising ET and 2CE; the B-DECIGO--ET network, which combines B-DECIGO and E  the B-DECIGO--2CE network, involving B-DECIGO and 2CE; and finally, the B-DECIGO--ET2CE network, incorporating B-DECIGO and ET2CE. 

\subsection{Calculation of SNR}\label{SNR}

When simulating GW detection, we set the detection threshold of SNR to 8 for both individual detectors and detector networks. This threshold is commonly used in the simulation of GW standard sirens, as it balances the need for sufficient detection confidence while maintaining a reasonable event rate \cite{LIGOScientific:2014pky,LIGOScientific:2017adf,LIGOScientific:2016fbo}.
The $\rm SNR$ of the detector network composed of $N$ GW detectors can be expressed as
\begin{equation}
\rho=\sqrt{\sum_{k=1}^{N}(\tilde{h}_{k}|\tilde{h}_{k})},
\end{equation}
where $\tilde{h}_{k}$ is the GW waveform of the $k$th GW detector. 
The inner product is defined as
\begin{equation}
(\tilde{h}|\tilde{h})=4\int_{f_{\rm in}}^{f_{\rm out}}\frac{\tilde{h}(f)\tilde{h}^{*}(f)}{S_{\rm n}(f)}\mathrm{d}f,
\end{equation}
where $\tilde{h}^{*}(f)$ is the complex conjugate of $\tilde{h}(f)$ and $S_{\rm n}(f)$ is the one-side noise power spectral density (PSD) of the GW detector. Here we adopt the PSD of ET from ref.~\cite{Hild:2010id}, of CE (40 km) from ref.~\cite{CE}, of CE (20 km) from ref.~\cite{Hild:2010id}, and of B-DECIGO from ref.~\cite{Isoyama:2018rjb}. Moreover, $f_{\rm in}$ and $f_{\rm out}$ are the frequencies at which the GW signal enters and leaves the frequency band of the GW detector, respectively.

\begin{figure}[!htbp]
\includegraphics[width=0.8\textwidth]{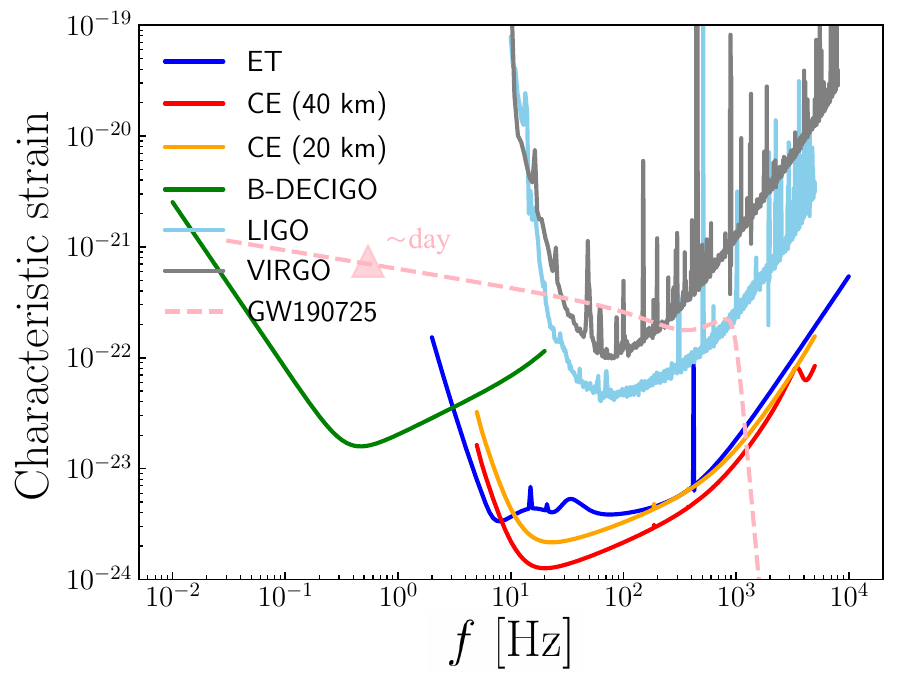}
\centering
\caption{Characteristic strains of ET, CE (40 km), CE (20 km), B-DECIGO, LIGO in Livingston, and VIRGO, together with the effective strain amplitude of GW190725. The pink triangle positions the frequency of the GW signal a day before coalescence. We define the dimensionless characteristic strain as $\sqrt{f{S}_{\rm n}}$ for the GW detectors and $2f|h(f)|$ for the GW source.
}\label{fig1}
\end{figure}

In figure~\ref{fig1}, we show the characteristic sensitivities of ET, CE (40 km), CE (20 km), B-DECIGO, LIGO in Livingston, and VIRGO (see ref.~\cite{KAGRA:2021vkt}). Note that the GW190725-like event is shown as the representative since it has the highest calculated SNR of the multi-band observation among the GW events.
It can be found that the frequency band of B-DECIGO is well connected with ground-based detectors, enabling the accumulation of a higher SNR across the longer frequency band.

\begin{figure}[!htbp]
\includegraphics[width=0.8\textwidth]{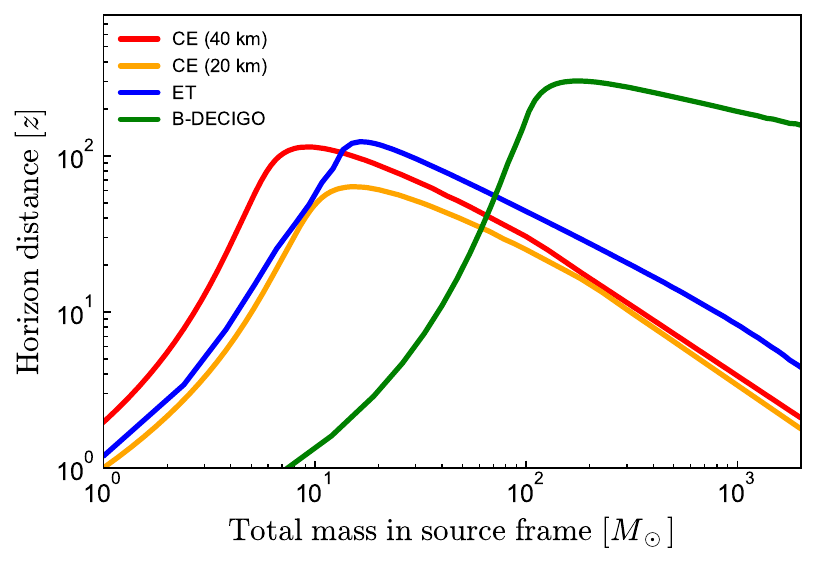}
\centering
\caption{Detection horizons for the equal-mass non-spinning binaries as a function of the source frame total mass for ET, CE (40 km), CE (20 km) and B-DECIGO.}
\label{fig:horizen}
\end{figure}

In figure~\ref{fig:horizen}, we show the corresponding detection horizons for equal-mass non-spinning binaries from ET, CE (40 km), CE (20 km), and B-DECIGO, based on the total mass in the source frame, with SNR $> 8$. We can find that, for B-DECIGO and 3G ground-based detectors, multi-band synergetic detection is achievable for both SMBHs and light intermediate-mass black holes.

\subsection{Fisher information matrix}\label{sec:FIM}
We use the Fisher information matrix (FIM) \cite{Finn:1992wt} to simulate the GW source parameters' measurement errors. For a GW detector network with $N$ interferometers, the FIM is given by
\begin{equation}
F_{ij}=\sum_{k=1}^{N}\left(\frac{\partial\tilde{h}_{k}}{\partial \theta_i}\bigg|\frac{\partial\tilde{h}_{k}}{\partial \theta_j}\right),
\end{equation}
where $\theta_i$ denotes the $i$th parameter among the nine GW source parameters describing the GW signal, which are $d_{\rm L}, t_{\rm c}, \mathcal M_{\rm c}, \eta, \theta, \phi, \psi, \iota, \psi_{\rm c}$. Among them, $\theta$ and $\phi$ represent the colatitude and the longitude of the GW event, respectively. $\psi$ is the polarization angle, $\psi_{\rm c}$ is the coalescence phase and $t_{\rm c}$ is the coalescence time.

The 9$\times$9 covariance matrix ($Cov$) of the GW source parameters is equal to the inverse of the FIM, and the measurement error of the $i$th GW parameter is given by $\Delta\theta_i=\sqrt{Cov_{ii}}$. 
The sky localization error is given as
\begin{equation}
\Delta\Omega = 2\pi|\sin{\theta}|\sqrt{(\Delta\theta)^2(\Delta\phi)^2 - (\Delta\theta\Delta\phi)^2}.
\end{equation}
The total error of $d_{\rm L}$ is expressed as
\begin{equation}\label{eq:sigma_dl}
\Delta d_{\rm L} = \sqrt{(\Delta d_{\rm L}^{\rm inst})^2 + (\Delta d_{\rm L}^{\rm lens})^2 + (\Delta d_{\rm L}^{\rm pv})^2},
\end{equation}
where $\Delta d_{\rm L}^{\rm inst}$ represents the instrumental error of $d_{\rm L}$, estimated using FIM; $\Delta d_{\rm L}^{\rm lens}$ denotes the weak-lensing error, given by \cite{Hirata:2010ba,Tamanini:2016zlh}
\begin{equation}
\Delta d_{\rm L}^{\rm lens}(z) = d_{\rm L}\times 0.066\left[\frac{1 - (1 + z)^{-0.25}}{0.25} \right]^{1.8}.
\end{equation}
$\Delta d_{\rm L}^{\rm pv}(z)$ is the peculiar-velocity error, given by \cite{Kocsis:2005vv}
\begin{equation}
\Delta d_{\rm L}^{\rm pv}(z) = d_{\rm L}\times \left[1 + \frac{c(1 + z)^2}{H(z)d_{\rm L}(z)} \right]\frac{\sqrt{\langle v^2 \rangle}}{c},
\end{equation}
where $\sqrt{\langle v^2 \rangle}$ is the peculiar velocity of the GW source, set to $\sqrt{\langle v^2 \rangle} = 500\ \rm km\ s^{-1}$ \cite{He:2019dhl}.

\subsection{Identifying GW events' potential host galaxies}\label{sec:identifying}

We rely on the three-dimensional (3D) localization capability of the GW detector to identify the potential host galaxies of GW sources. In our analysis, we model the 3D localization region of each GW event as a truncated cone, characterized by a radial range of $[d_{\rm L}^{\rm min},\ d_{\rm L}^{\rm max}] = [\bar{d}_{\rm L}-3\Delta d_{\rm L},\ \bar{d}_{\rm L}+3\Delta d_{\rm L}]$ and an angular region defined by $\chi^2\leq9.21$. Here, $\bar{d}_{\rm L}$ represents the luminosity distance of the GW event, $\Delta d_{\rm L}$ denotes the 1$\sigma$ error of the luminosity distance measurement, obtained by eq.~(\ref{eq:sigma_dl}), and $\chi^2$ is expressed as
\begin{equation}\label{eq:chi2}
    \begin{aligned}
        \chi^2=(\theta-\bar{\theta},\phi-\bar{\phi})Cov'^{-1}\begin{pmatrix}\theta-\bar{\theta}\\ \phi-\bar{\phi}
\end{pmatrix},
    \end{aligned}
\end{equation}
where $Cov'^{-1}$ is the $2\times2$ covariance matrix of $\theta$ and $\phi$. We obtain $Cov'^{-1}$ by removing the rows and columns associated with the other parameters from the original $9\times9$ covariance matrix estimated through the FIM analysis. The $\chi^2$ value quantifies the deviation of a data point at $(\theta,\phi)$ from the expected position at $(\bar{\theta},\bar{\phi})$ in terms of angle. The condition $\chi^2\leq9.21$ corresponds to the 99\% confidence region. 

\begin{figure}
    \centering
    \includegraphics[width=1\linewidth]{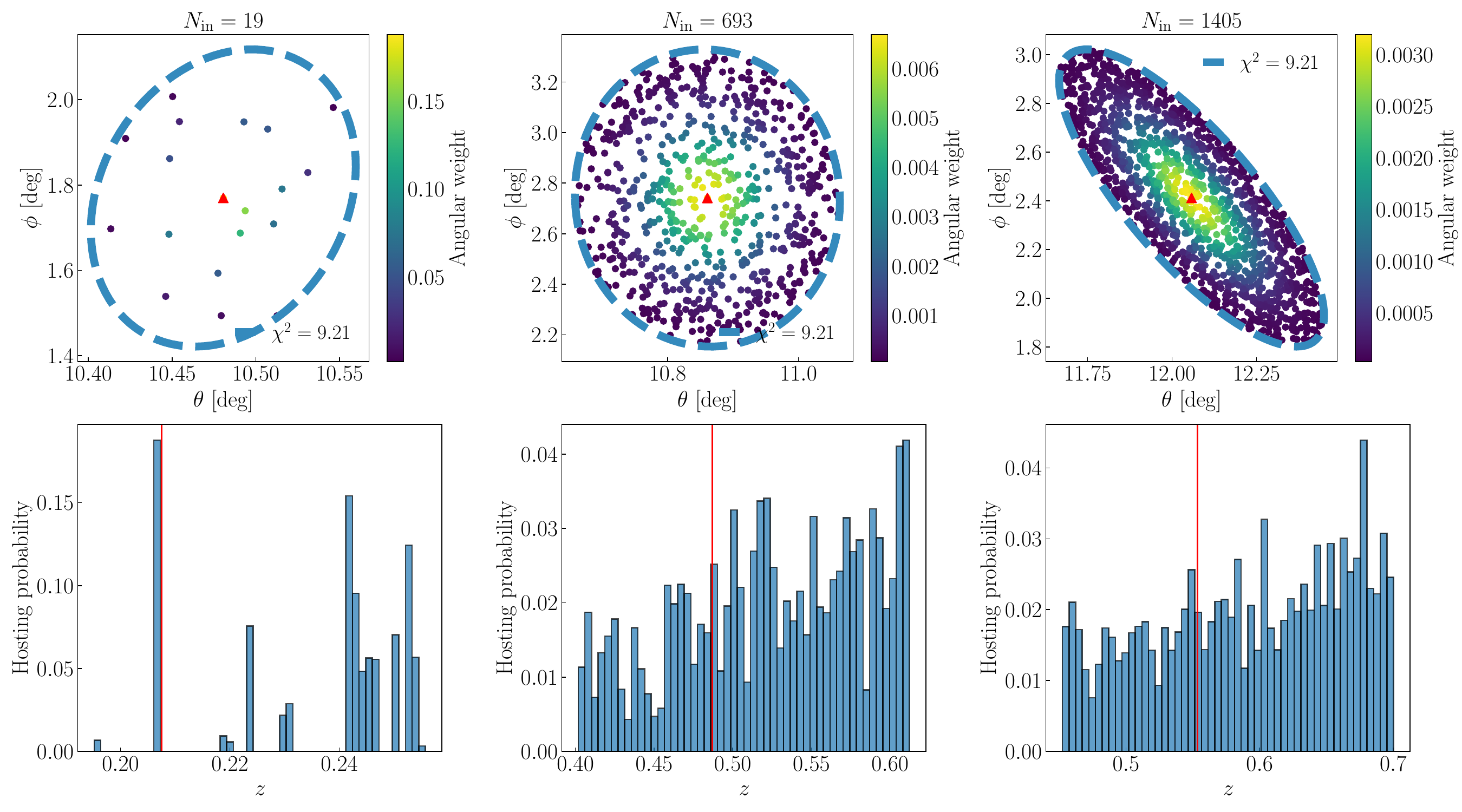}
    \caption{Process of identifying potential host galaxies, with each column showcasing a representative case. The upper panels depict the distribution of potential host galaxies in the $\theta$--$\phi$ plane. Each point is color-coded according to the angular weight of the potential host galaxy calculated using eq.~(\ref{eq:weight}). Red triangles denote the sky positions of the true host galaxies. The blue dashed circle represents the range of $\chi\leq9.21$, calculated through eq.~(\ref{eq:chi2}), corresponding to 99\% confidence region. The lower panels show the redshift distributions of potential host galaxies, with the y-axis indicating the probability of hosting GW sources in each redshift bin after using the angular weight. The red lines indicate the redshift of the true host galaxies.}
    \label{fig:identifying}
\end{figure}

To match with the galaxy catalog, we convert the range of luminosity distances into a corresponding range of redshifts, denoted as [$z_{\rm min}$,\ $z_{\rm max}$].  Specifically, $z^{\rm min}=z(d_{\rm L}^{\rm min}, H_0^{\rm min}, \Omega_{\rm m}^{\rm min})$ and $z^{\rm max}=z(d_{\rm L}^{\rm max}, H_0^{\rm max}, \Omega_{\rm m}^{\rm max})$, where $H_0^{\rm min}$, $H_0^{\rm max}$, $\Omega_{\rm m}^{\rm min}$, and $\Omega_{\rm m}^{\rm max}$ are the edge values of prior ranges of $H_0$ and $\Omega_{\rm m}$. Here, we set $H_0\in[20,\ 140]\ {\rm km\ s^{-1}\ Mpc^{-1}}$ and $\Omega_{\rm m}\in[0.1,\ 0.5]$. In summary, the potential host galaxies of each GW event are constrained within the range of [$z_{\rm min}$,\ $z_{\rm max}$] and $\chi^2\leq9.21$. 
Additionally, following ref.~\cite{Muttoni:2023prw}, we calculate an angular localization weight for each potential host galaxy using the equation
\begin{equation}
    \begin{aligned}
        w \propto \frac{1}{2\pi |Cov'|}\exp\left[-\frac{1}{2}(\theta-\bar{\theta},\phi-\bar{\phi})Cov'^{-1}\begin{pmatrix}\theta-\bar{\theta}\\ \phi-\bar{\phi}
\end{pmatrix}\right].
    \end{aligned}\label{eq:weight}
\end{equation}
For an intuitive description of the process of identifying potential host galaxies of GW sources, refer to figure~\ref{fig:identifying}.
It is worth noting that the angle weighting approach is only applicable in simulated data research. This is because the source parameter posterior in simulated data is assumed to be Gaussian, and the error is calculated through FIM. In real data research, however, the source parameter posterior is non-Gaussian, so angle weighting isn't feasible.

\subsection{Cosmological parameter inferences}\label{sec:cosmology inferrences}

Inferring cosmological parameters using GW dark sirens requires redshift information from galaxy catalogs. In this work, we generate a simplistic galaxy catalog by uniformly sampling galaxies within the comoving volume with a number density of 0.02 Mpc$^{-3}$ (the median value of the result of ref.~\cite{Barausse:2012fy}). We explore the impact of the galaxy catalog's completeness on cosmological parameter inference by changing the apparent magnitude threshold to mock galaxy catalogs with different levels of completeness. For redshift uncertainties of galaxies in the mock galaxy catalog, we ignore the redshift measurement errors for galaxies at $z\leq0.1$ assuming these galaxies could be covered by future spectroscopic surveys \cite{Gong:2019yxt,Song:2022siz}, and beyond this range, we mock the redshift uncertainties of galaxies as $\Delta z(z)=0.02(1+z)$, which is the redshift measurement accuracy could be achieved by future photometric surveys \cite{Cao:2017ph}.

We employ the Bayesian analysis to infer cosmological parameters $\bm{\Omega}$. The posterior distribution of $\bm{\Omega}$ is given by
\begin{equation}
    \begin{aligned}
        p(\bm{\Omega}|\{D_{\rm GW}\})\propto p(\{D_{\rm GW}\}|\bm{\Omega})p(\bm{\Omega}),
    \end{aligned}
\end{equation}
where $\{D_{\rm GW}\}$ represents the GW datase  $p(\bm{\Omega})$ is the prior distribution of $\bm{\Omega}$. We adopt uniform priors for all cosmological parameters, with $H_0\in$ [20,140] $\ksm$ and $\Omega_{\rm m}\in$ [0.1,0.5]. $p(\{D_{\rm GW}\}|\bm{\Omega})$ is the likelihood function. As the detection of each GW event is independent, the likelihood function of the GW data set $\{D_{\rm GW}\}$ can be expressed as the product of individual GW event's likelihood:
\begin{equation}
    \begin{aligned}
        p(\{D_{\rm GW}\}|\bm{\Omega})=\prod_{i=1}^{N_{\rm GW}}p(D_{{\rm GW},i}|\bm{\Omega}),
    \end{aligned}
\end{equation}
where $N_{\rm GW}$ is the number of GW events. The likelihood function of a single GW event is given by
\begin{equation}
    \begin{aligned}\label{eq:single gw events' likelihood}
        p(D_{\rm GW}|\bm{\Omega})=\frac{1}{\beta(\bm{\Omega})}\int\int p(D_{\rm GW}|d_{\rm L})\delta(d_{\rm L}-d_{\rm L}(z,\bm{\Omega}))p(z|\bm{\Omega}){\rm d}z{\rm d}d_{\rm L},
    \end{aligned}
\end{equation}
where $p(D_{\rm GW}|d_{\rm L})$ is the posterior distribution of the GW event's $d_{\rm L}$, given by
\begin{equation}
    \begin{aligned}
        p(D_{\rm GW}|d_{\rm L})=\frac{1}{\sqrt{2\pi}\Delta d_{\rm L}}\exp\left[-\frac{(\hat{d}_{\rm L}-d_{\rm L})^2}{2\Delta d_{\rm L}^2}\right],
    \end{aligned}
\end{equation}
where $\Delta d_{\rm L}$ is the 1-$\sigma$ error of $d_{\rm L}$ calculated using eq.~(\ref{eq:sigma_dl}), $\hat{d}_{\rm L}$ is the luminosity distance of the GW event. $d_{\rm L}(z,\bm{\Omega})$ is the theoretical luminosity distance calculated using eq.~(\ref{eq:d-z relation}) with $z$ and $\bm{\Omega}$. $p(z|\bm{\Omega})$ is the prior distribution of the GW source's redshift $z$ and is given by
\begin{equation}
    \begin{aligned}
        p(z|\bm{\Omega})=\bigg{\{}\frac{1}{N_{\rm in}}\sum^{N_{\rm in}}_{j=1}w_{j}\frac{1}{\sqrt{2\pi}\Delta z(\hat{z}_{j})}\exp\left[-\frac{(\hat{z}_{j}-z)^2}{2\big(\Delta z(\hat{z}_{j})\big)^2}\right] p({\rm G}|z,{\bm\Omega})+p({\rm \bar{G}}|z,\bm{\Omega})\bigg{\}}R_{\rm obs}(z),
    \end{aligned}
\end{equation}
where $\Delta z(\hat{z}_{j})$ is the redshift uncertainties of the $j$th potential host galaxy, $w_{j}$ is the angular weight of the $j$th potential host galaxy calculated using eq.~(\ref{eq:weight}), and $R_{\rm obs}(z)$ is the merger rate of SBBHs in the observer frame, obtained by eq.~(\ref{eq:Robs}). $p({\rm G}|z,{\bm\Omega})$ and $p({\rm \bar{G}}|z,\bm{\Omega})$ represent the completeness and incompleteness of the galaxy catalog at $z$, describing the probability of a galaxy at $z$ is in and is not in the galaxy catalog, respectively. We obtain $p({\rm G}|z,{\bm\Omega})$ from $p({\rm G}|d_{\rm L})$ through the distance-redshift relation, and $p({\rm \bar{G}}|z,\bm{\Omega}) = 1-p({\rm G}|z,{\bm\Omega})$.

Limited by the observation capability of the survey telescope, the distant and faint galaxy may be ignored by the galaxy catalog, causing the completeness of the galaxy catalog to drop with the distance. To estimate $P({\rm G}|d_{\rm L},{\bm\Omega})$, we first mock galaxies' luminosities by assuming the luminosity distribution of galaxies in the universe can be described by the Schechter function \cite{Schechter:1976iz}, given by
\begin{equation}
    p(L)\propto \left(L / L^{*}\right)^{\alpha} \exp \left(-L / L^{*}\right) \mathrm{d} L / L^{*},
\end{equation}
where $p(L)$ is the probability of galaxies with luminosity $L$, $\alpha=-1.07$, and $L^* = 1.2\times10^{10}h^{-2}L_{\odot}$ is the characteristic galaxy luminosity, with $L_{\odot}$ being the solar luminosity. Following ref.~\cite{Gray:2019ksv}, we set the lower luminosity cutoff for the dimmest galaxies in the universe to $0.001L^*$. The luminosity is then converted to the apparent magnitude, ${mag}$, via 
\begin{equation}
    {mag}\left(L, d_{\rm L}\right)= Mag_{\odot}-2.5 \log _{10}^{\left(L / L_{\odot}\right)}+5 \log _{10}^{\left(d_{\rm L} / \mathrm{pc}\right)}-5,
\end{equation}
where $Mag_{\odot}$ is the absolute magnitude of the sun. We estimate $P({\rm G}|d_{\rm L},{\bm\Omega})$ by calculating the fraction of galaxies' apparent magnitudes lower than the apparent magnitude threshold in the $d_{\rm L}$ bins with a bin width of 17 Mpc.

Noticeably, in the real observed galaxy catalog, the completeness may vary with the sky localization, so we should calculate $P({\rm G}|d_{\rm L},{\bm\Omega})$ pixel by pixel \cite{Gray:2019ksv}. In our work, we adopt a simplified assumption that the completeness of the galaxy catalog is uniform in the sky.

The GW selection effect $\beta(\bm{\Omega})$, as delineated in ref.~\cite{Chen:2017rfc}, is considered, expressed as:
\begin{equation}
    \begin{aligned}
        \beta(\bm{\Omega})=\int p_{\rm det}^{\rm GW}\big(d_{\rm L}(z,\bm{\Omega})\big)p(z|\rm{\Omega}){\rm d}z,
    \end{aligned}
\end{equation}
where $p_{\rm det}^{\rm GW}\big(d_{\rm L}(z,\bm{\Omega})\big)$ represents the detection probability of the GW event at $d_{\rm L}(z,\bm{\Omega})$, which can be evaluated using Monte-Carlo integration, as explicated in ref.~\cite{Gray:2019ksv},
\begin{equation}
    \begin{aligned}
        p_{\rm det}^{\rm GW}(d_{\rm L})=\frac{1}{N_{\rm samp}}\sum_{n=1}^{N_{\rm samp}}p_{\rm det}^{\rm GW}(d_{\rm L}|\bm{\theta}_{n}),
    \end{aligned}
\end{equation}
with
\begin{equation}
    \begin{aligned}
        p_{\rm det}^{\rm GW}(d_{\rm L}|\bm{\theta}_n)\approx\left\{\begin{array}{ll}
1, & \text { if } \rho_{n}>\rho_{\rm th}, \\
0, & \text { otherwise, }
\end{array}\right.
    \end{aligned}
\end{equation}
where $N_{\rm samp}=50000$ represents the total number of Monte-Carlo realizations, and $\bm{\theta}_{n}$ are the GW source parameters other than $d_{\rm L}$ and $z$ in the $n$th sampling. We conduct 50000 simulations at each 50 Mpc interval of $d_{\rm L}$. When simulating, the source parameters, excluding $d_{\rm L}$ and $z$, are drawn from the population models outlined in section~\ref{sec:simulate GW source}. Ultimately, we derive a smooth curve describing the distribution of $p_{\rm det}^{\rm GW}$ along $d_{\rm L}$, free from any fluctuations or jitter.

\section{Results and discussion}\label{sec:results and discussion}

In this section, we show the results of our analysis, including localization errors of GW events and constraints on cosmological parameters, and make relevant discussions.

\subsection{Localizations of SBBHs}

\begin{figure*}[!htbp]
\includegraphics[width=0.95\textwidth]{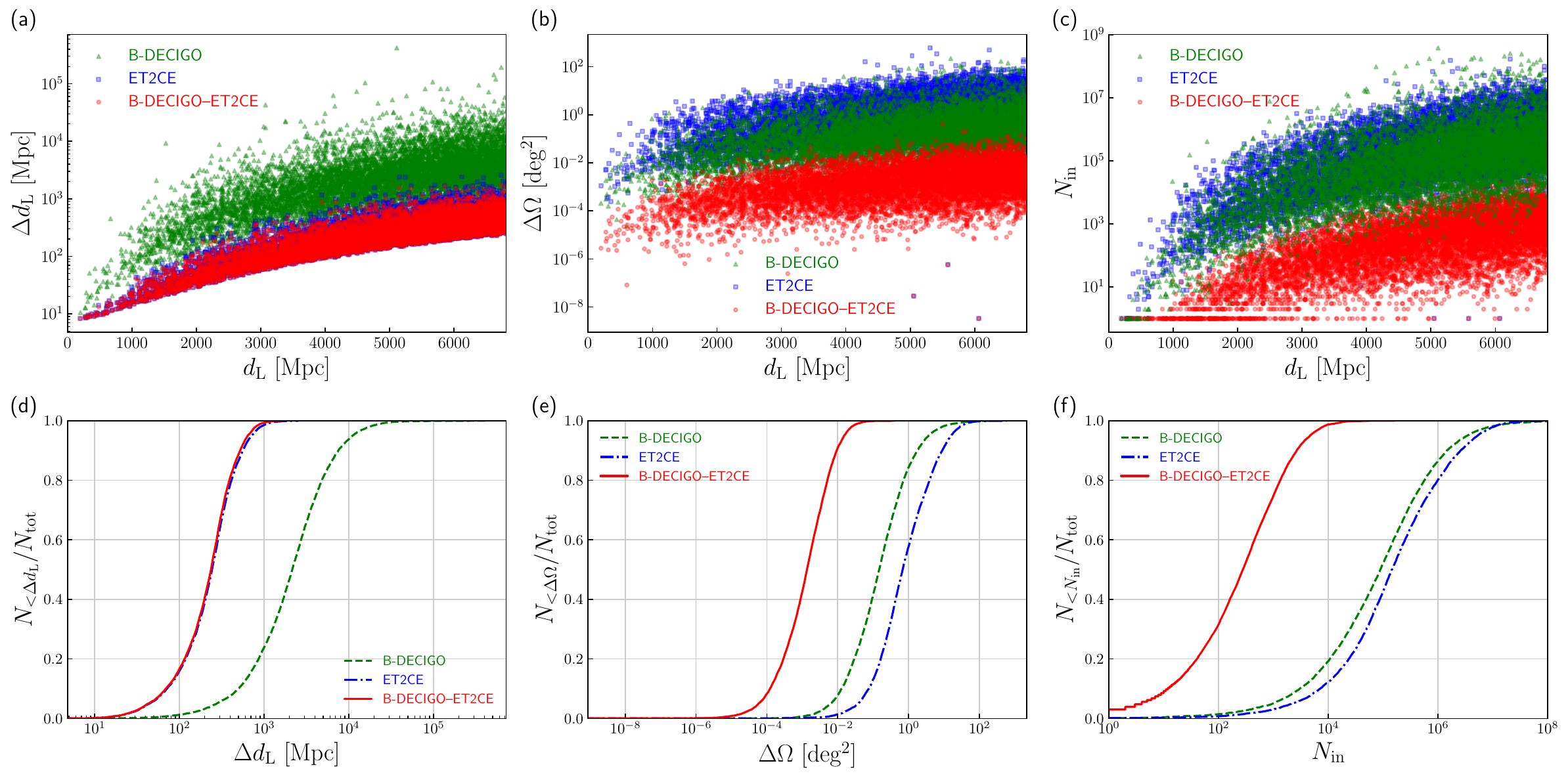}
\centering
\caption{The localization error and the number of potential host galaxies of mock GW events. Panel (a): The results of $\Delta d_{\rm L}$ across four GW detector configurations: ET2CE network, B-DECIGO, B-DECIGO--ET2CE network, and LVK network. Panel (b): Similar to panel (a), but showcasing the results for $\Delta\Omega$. Panel (c): Similar to panel (a), but showcasing the number of potential host galaxies $N_{\rm in}$ of each GW event. Panels (d)--(f): CDF of GW events on $\Delta d_{\rm L}$, $\Delta\Omega$, and $N_{\rm in}$, respectively. In the CDF curve of  $\Delta d_{\rm L}$, the y-axis value corresponding to an x-axis value of $X$ represents the proportion of GW events with $\Delta d_{\rm L}$ less than $X$ relative to the total number of events, and similarly for the CDF curves of $\Delta\Omega$ and $N_{\rm in}$.}\label{fig:errors}
\end{figure*}

We show the scatter distributions and cumulative distribution functions (CDFs) of $\Delta d_{\rm L}$, $\Delta \Omega$, and $N_{\rm in}$ in figure~\ref{fig:errors}. In determining $d_{\rm L}$, we find that B-DECIGO--ET2CE performs similarly to ET2CE, outperforming B-DECIGO with $\Delta d_{\rm L}$ being approximately one order of magnitude smaller. This occurrence is because the measurement of $d_{\rm L}$ depends on the SNR of the detected signal. The signals of SBBH mainly fall within the detection band of ground-based detectors. Therefore, the measurement error of $d_{\rm L}$ for ET2CE is smaller than that for B-DECIGO. Moreover, the network composed of B-DECIGO and ET2CE does not show significant enhancement in determining $d_{\rm L}$ compared to ET2CE alone. For spatial localization, the spatial localization of B-DECIGO--ET2CE is two orders of magnitude better than that of B-DECIGO and ET2CE, and B-DECIGO is nearly an order of magnitude better than ET2CE.
The ability of GW detectors to measure $d_{\rm L}$ and perform spatial angular localization directly influences their search for potential host galaxies. We find that multi-band networked observations can simultaneously enhance the capabilities of measuring $d_{\rm L}$ and spatial localization, significantly reducing $N_{\rm in}$. The $N_{\rm in}$ distributions of B-DECIGO and ET2CE are similar, with B-DECIGO performing slightly better, mainly due to its superior spatial localization.

The multi-band synergetic observation of B-DECIGO and ET2CE can significantly improve spatial localization for GW events, which is mainly due to: (i) The multi-band synergetic observation can significantly improve SNR; (ii) The multi-band synergetic observation allows for long-time observation of the GW signal. During this period, the GW detector's orbital motion enhances spatial positioning accuracy; (iii) The network of space-based and ground-based GW detectors has long baselines, directly boosting the spatial localization.

\subsection{Constraints on cosmological parameters}

We marginalize $d_{\rm L}$ and $z$ to infer cosmological parameters. We consider the merger rate distribution discussed in section~\ref{sec:simulate GW source} as the population model for GW sources in the redshift prior distribution and assume that we perfectly know the values of population model parameters. 
In the analysis based on real GW data, a joint inference of cosmological parameters and population model parameters is a better approach \cite{Mastrogiovanni:2023emh,Gray:2023wgj}, due to the poor constraints on population model parameters from current GW observations, and some of these parameters may degenerate with cosmological parameters \cite{LIGOScientific:2021aug}.

We only consider GW events at $z\leq1$, as including larger redshift ranges will drastically increase the number of GW events, leading to excessively long computation times. In fact, the multi-band joint detection of ET2CE and B-DECIGO can detect nearly all GW sources predicted by the population model, with redshifts extending up to $z\sim10$. Additionally, future galaxy surveys may cover redshift ranges as high as 4 \cite{Gong:2019yxt}. Therefore, incorporating GW sources with $z>1$ could further enhance the constraint constraints of cosmological parameters predicted in this study. One potential solution to the issue of excessive computational cost is to use deep learning as an alternative to Bayesian inference \cite{Stachurski:2023ntw}. However, this approach falls outside the scope of our current research and may be considered in future studies.

\begin{figure}[!htbp]
\includegraphics[width=0.8\textwidth]{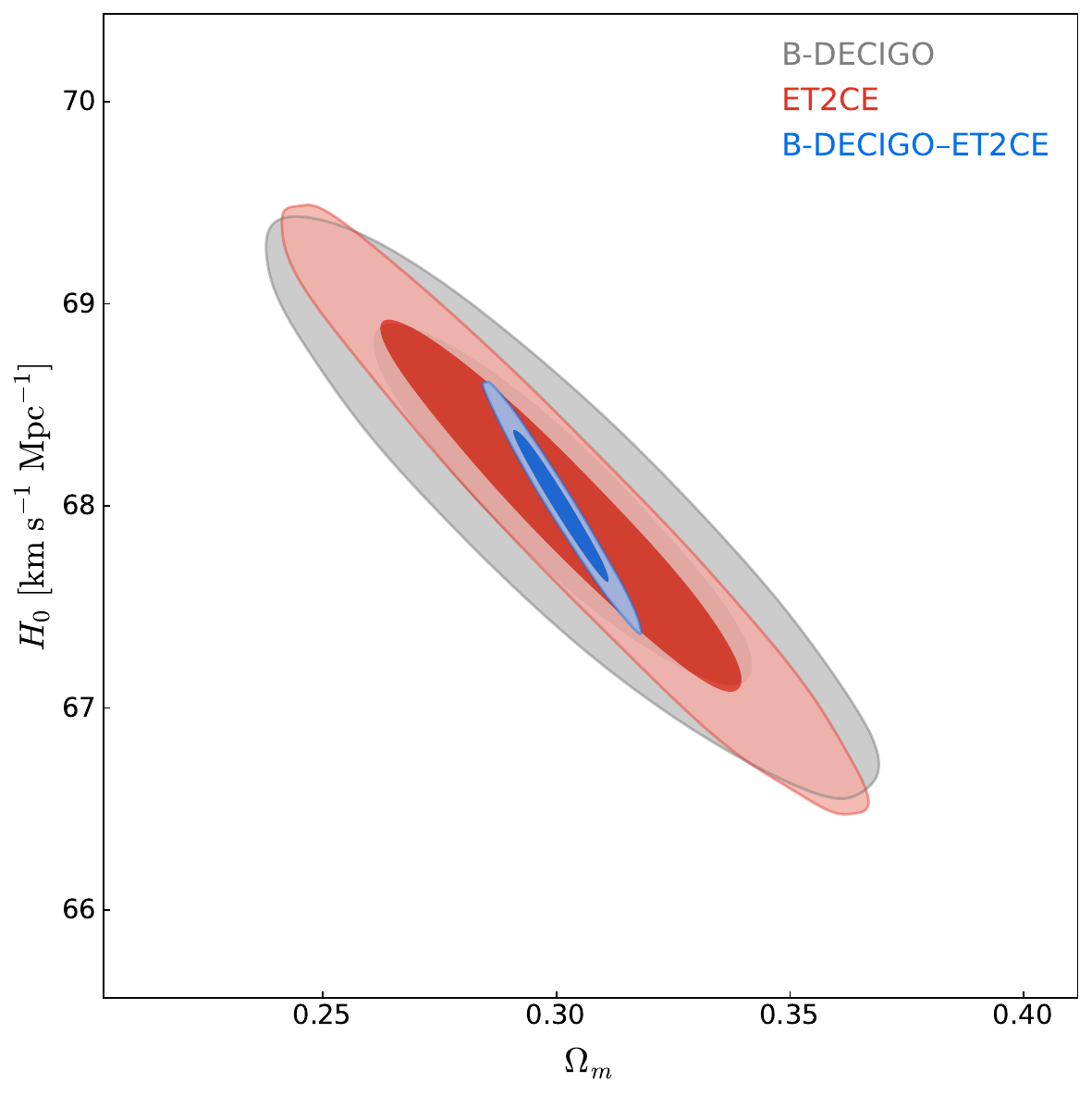}
\centering
\caption{Two-dimensional marginalized contours ($68.3\%$ and $95.4\%$ confidence level) in the $\Omega_{\rm m}$--$H_0$ plane by using B-DECIGO, ET2CE and B-DECIGO--ET2CE mock data for the $\Lambda$CDM model.}\label{fig:contour}
\end{figure}

\begin{figure}[!htbp]
\includegraphics[width=0.8\textwidth]{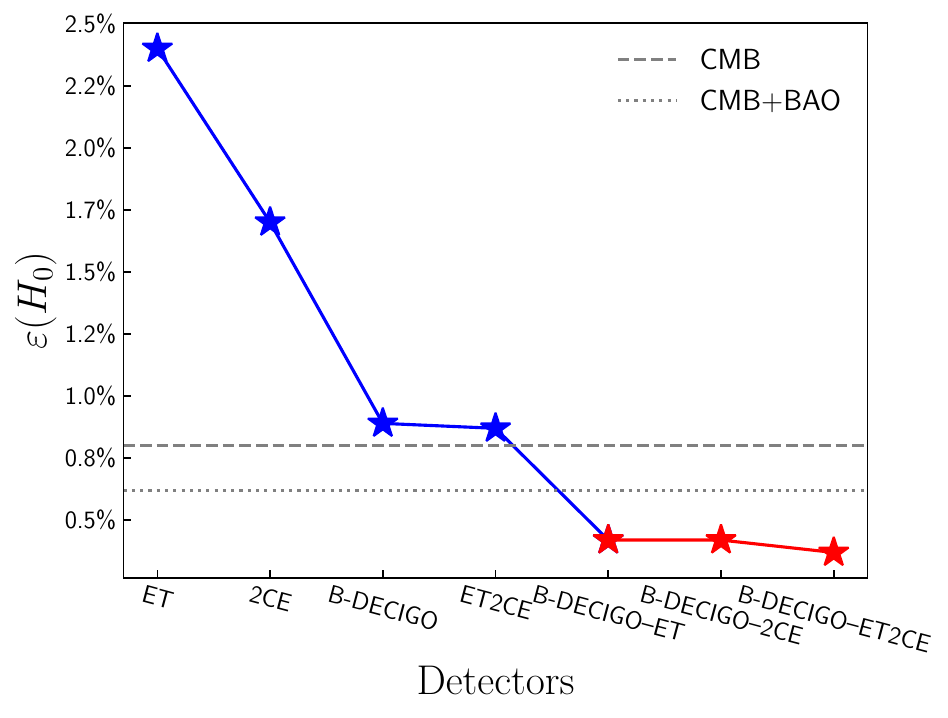}
\centering
\caption{The relative errors of $H_{0}$ by using ET, 2CE, B-DECIGO, ET2CE, B-DECIGO--ET, B-DECIGO--2CE, and B-DECIGO--ET2CE mock data. The gray horizontal dashed and point lines represent the constraint precisions of Planck2018 CMB and CMB+BAO data.}\label{fig:H0 constraints}
\end{figure}

\begin{figure}[!htbp]
\includegraphics[width=0.8\textwidth]{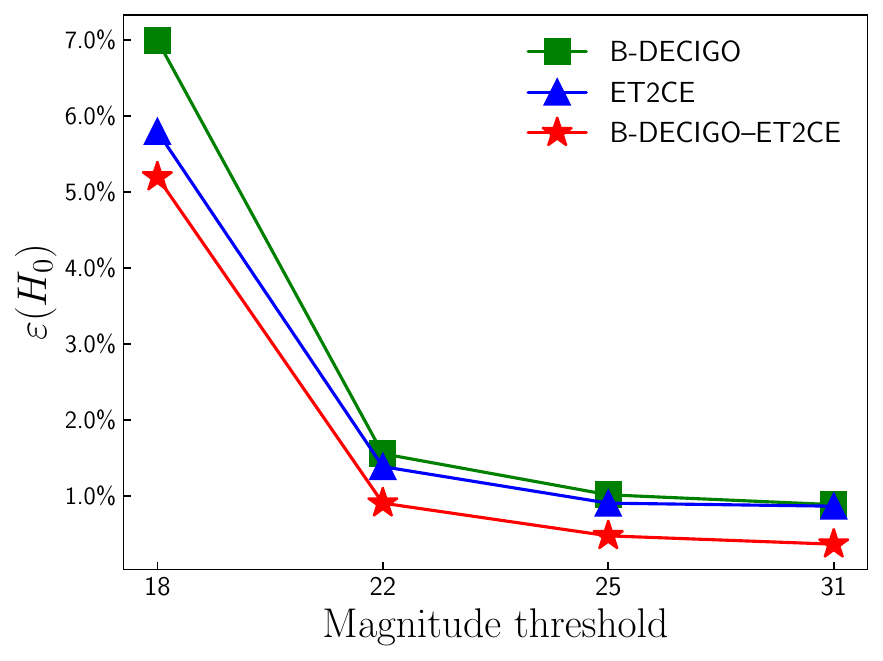}
\centering
\caption{The relative errors of $H_{0}$ for different magnitude thresholds by using B-DECIGO, ET2CE and B-DECIGO--ET2CE.}\label{fig:H0mag}
\end{figure}

In this study, we investigate the impact of varying levels of catalog completeness by adjusting the apparent magnitude threshold of the mock galaxy catalog to values of 18, 22, 25, and 31. Here, a magnitude threshold of 31 represents a galaxy catalog with 100\% completeness at $z\leq1$. The results presented in table~\ref{tab1}, figure~\ref{fig:contour}, and figure~\ref{fig:H0 constraints} are derived using the mock galaxy catalog with a magnitude threshold of 31. Furthermore, we quantify the dependence of the $H_0$ constraint precision on the magnitude threshold in figure~\ref{fig:H0mag}, providing insights into how catalog completeness affects cosmological parameter estimation.

Table~\ref{tab1} summarizes the results obtained using a galaxy catalog with 100\% completeness at $z\leq1$. In the first rows, we show the number of GW events with $N_{\rm in}=1$. GW events with $N_{\rm in} = 1$ can be assumed as bright sirens, with redshifts determined by the follow-up spectroscopic observation. Using the multi-band GW synergetic detection can significantly increase the number of GW events with $N_{\rm in} = 1$. In the second to fifth rows, we report parameter constraints for $H_0$ and $\Omega_{\rm m}$ in the $\Lambda$CDM model.

\begin{table*}[!htbp]
\renewcommand\arraystretch{1.5}
\centering
\vspace{2mm}
\normalsize
\setlength{\tabcolsep}{3mm}{
\resizebox{\textwidth}{!}{
\begin{tabular}
{p{1.5cm}<{\centering} p{1.3cm}<{\centering} p{1.3cm}<{\centering} p{2.0cm}<{\centering} p{1.5cm}<{\centering} p{2.2cm}<{\centering} p{2.2cm}<{\centering} p{2.2cm}<{\centering}}
\hline\hline
Result type & ET & 2CE & B-DECIGO & ET2CE & B-DECIGO--ET & B-DECIGO--2CE & B-DECIGO--ET2CE \\
\hline
$N_{\rm in}=1$ & 2 & 5 & 15 & 13 & 244 & 257 & 276\\
\hline
$\sigma(\Omega_{\rm m})$& 0.083 & 0.057 & 0.030 & 0.026 & 0.008 & 0.008 & 0.007 \\
$\sigma(H_0)$   & 1.62 & 1.12 & 0.62 & 0.61 & 0.30 & 0.29 & 0.26\\
$\varepsilon(\Omega_{\rm m})$ & $29.5\%$ & $18.8\%$ & $9.8\%$ & $8.4\%$ & $3.4\%$ & $3.4\%$ & $3.0\%$ \\
$\varepsilon(H_0)$  &$2.41\%$ & $1.79\%$ & $0.89\%$ & $0.87\%$ & $0.42\%$ & $0.42\%$ & $0.37\%$  \\
\hline\hline
\end{tabular}
}}
\caption{
The number of GW events with $N_{\rm in}=1$, alongside the absolute errors (1$\sigma$) and the relative errors of the cosmological parameters in the $\Lambda$CDM model. Here the unit of $H_0$ is $\rm km\ s^{-1}\ Mpc^{-1}$.}\label{tab1}
\end{table*}

In figure.~\ref{fig:contour}, we present the constraint results of B-DECIGO, ET2CE, and B-DECIGO--ET2CE in the $H_0$--$\Omega$ plane, using a galaxy catalog with 100\% completeness at $z\leq1$. They depict the optimal situations of the space-based and ground-based detectors and the multi-band GW synergetic detector network, respectively.
Using the multi-band GW synergetic detection, B-DECIGO--ET2CE outperforms ET2CE and B-DECIGO, exhibiting enhanced constraint precisions on $H_0$ by 57.5\% and 58.4\% and on $\Omega_{\rm m}$ by 64.2\% and 69.4\%, respectively. 
Despite B-DECIGO--ET2CE shares a similar $\Delta d_{\rm L}$ distribution with ET2CE, as seen in panels (a) and (d) of figure.~\ref{fig:errors}, its lower $\Delta \Omega$ leads to the smaller $N_{\rm in}$, indicating more precise redshift inference than ET2CE, which enhances precision in cosmological parameter constraints through the distance-redshift relation. Compared with B-DECIGO, as shown in figure.~\ref{fig:errors}, B-DECIGO--ET2CE has advantages in both $\Delta d_{\rm L}$ and $\Delta \Omega$, leading to tighter constraints on cosmological parameters.

In figure~\ref{fig:H0 constraints}, we present the $H_0$ constraint precisons of all GW detector scenarios, based on a galaxy catalog with 100\% completeness at $z\leq1$. 2CE possesses an enhancement of 25.7\% over ET.
The enhancement of B-DECIGO compared with 2CE is significant, with an $H_0$ constraint precision enhancement of 50.3\%. When using the multi-band GW synergetic detection, the precision constraint precisions of $H_0$ are below 0.5\%, better than the Planck 2018 CMB+BAO result \cite{Planck:2018vyg}. Even in the worst case of using multi-band GW synergetic detection, B-DECIGO--ET significantly outperforms the best case of not using (ET2CE), with the $H_0$ constraint precision improved by 51.7\%. Notably, with an increasing number of ground-based GW detectors in the multi-band GW synergetic detector network, there are only slight improvements in $H_0$ constraint precisions. Specifically, with two decimal places retained, the enhancement of B-DECIGO--2CE over B-DECIGO--ET is not visible, and B-DECIGO--ET2CE exhibits a 11.9\% improvement over B-DECIGO--2CE. A similar situation also appears in constraining $\Omega_{\rm m}$, as shown in table~\ref{tab1}. The $\Omega_{\rm m}$ constraint precision of B-DECIGO--ET is improved by 59.5\% compared with that of ET2CE. Meanwhile, the enhancement of B-DECIGO--2CE over B-DECIGO--ET is also not visible, and B-DECIGO--ET2CE is 11.8\% better than B-DECIGO--2CE. Our results indicate that the multi-band GW synergetic observation can significantly enhance the constraint precisions of cosmological parameters and may help resolve the Hubble tension.

In figure~\ref{fig:H0mag}, we demonstrate the variation in $H_0$ constraint precisions when adopting different apparent magnitude thresholds. We find that the $H_0$ constraint precisions decrease as the apparent magnitude threshold decreases and the galaxy catalog's completeness declines. The thresholds of 18 and 22 roughly represent the upper limits of apparent magnitudes for current survey telescopes, and the threshold of 25 roughly represent the upper limits of apparent magnitudes for future survey telescopes, such as LSST \cite{LSST:2008ijt,LSSTScience:2009jmu} and CSST \cite{Cao:2017ph,Cao:2021bqm}. From our results, it is evident that the $H_0$ constraint precisions can be significantly improved with the deployment of future survey projects. When the next-generation survey telescopes become operational, using GW dark sirens could enable an around 1\% constraint precision for $H_0$ and even a sub-precision measurement of $H_0$ with future GW detector networks, which will provide substantial assistance in resolving the Hubble tension.

\section{Conclusion}\label{Conclusion}

GW standard sirens have significant potential as a cosmological probe for measuring absolute distances and can constrain the cosmic expansion history through the distance-redshift relation. The statistical galaxy catalog method provides redshifts for dark sirens without electromagnetic counterparts, relying on GW localization precisions. Multi-band GW synergetic detection can improve SNRs and localization precisions. In this paper, we explore how the multi-band GW synergetic detection of B-DECIGO and 3G ground-based GW detectors enhances dark siren cosmology.

Firstly, we simulate the detection of future GW detectors of SBBH events at $z\leq1$ in one year using the IMRPhenomD waveform model. Secondly, we use FIM to estimate the localization precisions of B-DECIGO, ET, CEs, and the GW detector networks composed of them. Thirdly, we identify potential host galaxies of dark sirens by cross-matching the GW localization region with a mock galaxy catalog complete up to $z = 1$. We also consider the influence of the completeness of the mock galaxy catalog on our cosmological inference results by setting different apparent magnitude thresholds. Finally, we use Bayesian analysis to infer cosmological parameters in the $\Lambda$CDM model.

Our results show that with one year of simulated SMBH GW events, the multi-band GW synergetic observation from B-DECIGO and 3G ground-based GW detectors can constrain $H_0$ to around 0.4\%, showing significant potential in dark siren cosmology. In comparison, without the multi-band synergetic observation strategy, B-DECIGO and 3G ground-based GW detectors can only constrain $H_0$ to 0.87\%--2.41\%. Using the multi-band synergetic detection strategy has more than 51\% improvement in the constraint precision on $H_0$ compared with not using it.

We find that multi-band GW synergetic observation enhances dark siren cosmology in the following aspects: (1) Extending the detection frequency range to improve the SNR, helping us to include more GW events in the dark siren analysis. (2) Enhancing the GW localization precisions, reducing the number of potential host galaxies, and helping to infer the redshifts of dark sirens. (3) Increasing the number of dark sirens with $N_{\rm in} = 1$, which are de facto bright sirens. We conclude that the multi-band GW synergetic observation has a significant potential enhancement to dark siren cosmology.

\section*{Acknowledgements}
This work was supported by the National SKA Program of China (Grants Nos. 2022SKA0110200 and 2022SKA0110203), the National Natural Science Foundation of China (Grants Nos. 11975072, 11875102, and 11835009), the National 111 Project (Grant No. B16009), and the China Scholarship Council.

\bibliography{multi-band}{}
\bibliographystyle{JHEP}

\end{document}